\documentclass{article}
\usepackage{amssymb,amsmath}
\numberwithin{equation}{section}
\newtheorem{lemma}{Lemma}[section]

\newtheorem{proposition}[lemma]{Proposition}
\newtheorem{theorem}[lemma]{Theorem}
\newtheorem{remark}[lemma]{Remark}

\begin{document}
\title{Global existence and future asymptotic behaviour for solutions of the
Einstein-Vlasov-scalar field system with surface symmetry}
\author{D. Tegankong \\
Department of Mathematics, ENS,\\
University of Yaounde 1, Box 47, Yaounde, Cameroon \\
{dtegankong@yahoo.fr} }
\date{}
\maketitle
\begin{abstract}
We prove in the cases of plane and hyperbolic symmetries a
 global in time existence result in the future for
cosmological solutions of the Einstein-Vlasov-scalar field system,
with the sources generated by a distribution function and a scalar
field, subject to the Vlasov and wave equations respectively.
 The spacetime is future geodesically complete in the special case of plane symmetry with only a scalar
 field.
Causal geodesics are also shown to be future complete for general
homogeneous solutions of the Einstein-Vlasov-scalar field system
with plane and hyperbolic symmetry.
\end{abstract}
\section{Introduction}
Consider the Einstein-Vlasov-scalar field system with spherical, plane and hyperbolic symmetries.
 For more information on
this system which describes the evolution of self-gravitating
collisionless matter and scalar waves within the context of
general relativity, see \cite{tegankong}. We study in this paper
global existence for solutions of this system where the Einstein
equations are coupled to the Vlasov and wave equations. This
completes in the expanding direction, the results of
\cite{tegankong} which concentrated on the local in time existence
results and also in the global existence in the special case of
plane symmetric Einstein-scalar field system (i.e without Vlasov
contribution). It is shown that in the spherical case, no solution
exist globally in the future. Using the results of \cite{jurke}
which studied the future asymptotics of polarized Gowdy
$T^3$-models, we prove in the plane symmetric case without Vlasov
that the spacetime obtained is geodesically complete in the
future. The same result holds for spatially homogeneous solutions
of the full system with plane or hyperbolic symmetry. Only a
linear scalar field is treated but it is likely that the approach
developed here will be useful in the nonlinear case.

 Let us recall the formulation of the Einstein-Vlasov-scalar field system as shown in \cite{tegankong}.
  We consider a four-dimensional spacetime manifold $M$, with local
coordinates $(x^{\alpha})= (t,x^{i})$ on which $x^{0}=t$ denotes
the time and $(x^{i})$ the space coordinates. Greek indices always
run from $0$ to $3$, and Latin ones from $1$ to $3$. On $M$, a
Lorentzian metric $g$ is given with signature $(-,+,+,+)$.
 We consider a self-gravitating collisionless gas and restrict ourselves to the
case where all particles have the same rest mass, normalized to
$1$, and move forward in time. We denote by $(p^{\alpha})$ the
momenta of the particles. The conservation of the quantity
$g_{\alpha\beta}p^{\alpha}p^{\beta}$ requires that the phase space
of the particle is the seven-dimensional submanifold
\begin{equation*}
     PM = \{g_{\alpha\beta}p^{\alpha}p^{\beta} = -1;\ \
     p^{0}>0\}
\end{equation*}
 of $TM$ which is coordinatized by $(t,x^{i}, p^{i})$. If the coordinates
are such that the components $g_{0i}$ vanish then the component
$p^{0}$ is expressed by the other coordinates via
\begin{equation*}
 p^{0} = \sqrt{-g^{00}}\sqrt{1+g_{ij}p^{i}p^{j}}
 \end{equation*}
The distribution function of the particles is a non-negative
real-valued function denoted by $f$, that is defined on $PM$. In
addition we consider a  massless scalar field $\phi$ which is a real-valued
function on $M$. The Einstein-Vlasov-scalar field system now
reads:
\begin{equation*}
\partial_{t}f + \frac{p^{i}}{p^{0}}\partial_{x^{i}}f -
\frac{1}{p^{0}}\Gamma_{\beta\gamma}^{i}p^{\beta}p^{\gamma}\partial_{p^{i}}f
 = 0
 \end{equation*}
\begin{equation*}
\nabla^\alpha\nabla_\alpha\phi=0
\end{equation*}
 \begin{equation*}
G_{\alpha\beta}  =  8 \pi T_{\alpha\beta}
\end{equation*}
\begin{equation*}
T_{\alpha\beta}  = -\int_{\mathbb{R}^{3}}fp_{\alpha}p_{\beta}\mid
g \mid^{\frac{1}{2}} \frac{dp^{1}dp^{2}dp^{3}}{p_{0}} +
(\nabla_{\alpha}\phi\nabla_{\beta}\phi -
\frac{1}{2}g_{\alpha\beta}\nabla_{\nu}\phi\nabla^{\nu}\phi)
\end{equation*}
where $p_{\alpha} = g_{\alpha\beta}p^{\beta}$,
 $|g|$ denotes the modulus of determinant of the metric $g_{\alpha\beta}$,
 $\Gamma_{\alpha\beta}^{\lambda}$
 the Christoffel symbols,
$G_{\alpha\beta}$ the Einstein tensor, and $T_{\alpha\beta}$ the
energy-momentum tensor.

Note that since the contribution of $f$ to the energy-momentum
tensor is divergence-free \cite {ehlers}, the form of the
contribution of the scalar field to the energy-momentum tensor
determines the field equation for $\phi$.

  We refer to \cite{rendall} for the notion of spherical, plane and hyperbolic  symmetry. We
now consider a solution of the Einstein-Vlasov-scalar field system
where all unknowns are invariant under one of these symmetries. We
write the system in areal coordinates, i.e coordinates are chosen
such that $R=t$, where $R$ is the area radius function on a
surface of symmetry. The
 circumstances under which coordinates of this type exist are
 discussed in \cite{andreasson}. In such coordinates the metric $g$ takes the form
\begin{equation} \label{eq:1.1}
ds^{2} = - e^{2\mu(t,r)}dt^{2} + e^{2\lambda(t,r)}dr^{2} +
t^{2}(d\theta^{2} + \sin_{k}^{2}\theta d\varphi^{2})
\end{equation}
where
\begin{equation*}
 \sin_{k}\theta =
\begin{cases}
  \sin \theta  &\text{for $k = 1$  (spherical symmetry);}  \\
  1           &\text{for $k = 0$  (plane symmetry);}  \\
  \sinh\theta  &\text{for $k = -1$  (hyperbolic symmetry)}
\end{cases}
\end{equation*}
 $t > 0$ denotes a time-like coordinate, $r\in \mathbb{R}$ and
$(\theta, \varphi)$ range
 in the domains $[0,\pi]\times[0,2\pi]$, $[0,2\pi]\times[0,2\pi]$,
$[0,\infty[\times[0,2\pi]$
 respectively, and stand for angular coordinates. The functions $\lambda$
and $\mu$ are periodic in $r$ with period
 $1$. It has been shown in \cite{rein1} that due
 to the symmetry, $f$ can be written as a function of
 \begin{equation*}
 t, r, w := e^{\lambda}p^{1} \  {\rm and} \  F := t^{4}[(p^{2})^{2} +
\sin_k^2\theta (p^{3})^{2}],
 \end{equation*}
i.e. $f = f(t,r,w,F)$. In these variables, we have $p^{0} =
e^{-\mu} \sqrt{1+ w^{2}+ F/t^{2}}$. The scalar field is a function
of $t$ and $r$ which is periodic in $r$ with period 1.

We denote by a dot and by a prime the derivatives of the metric
components and of the scalar field with respect to $t$ and $r$
respectively. Using the results of
\cite{tegankong}, the complete Einstein-Vlasov-scalar
field system can be written as follows:
\begin{equation} \label{eq:1.2}
\partial_{t}f + \frac{e^{\mu - \lambda}w}{\sqrt{1 + w^{2} +
F/t^{2}}} \partial_{r}f - (\dot{\lambda}w + e^{\mu - \lambda}
\mu'{\sqrt{1 + w^{2} + F/t^{2}}}) \partial _{w}f = 0
\end{equation}
\begin{equation} \label{eq:1.3}
e^{-2 \mu}(2t \dot{\lambda} + 1) + k = 8 \pi t^{2} \rho
\end{equation}
\begin{equation} \label{eq:1.4}
e^{-2\mu}(2t \dot{\mu} - 1) - k = 8 \pi t^{2}p
\end{equation}
\begin{equation} \label{eq:1.5}
\mu' = -4\pi t e^{\lambda + \mu}j
\end{equation}
\begin{equation} \label{eq:1.6}
e^{-2 \lambda}(\mu'' + \mu'(\mu' - \lambda')) - e^{-2
\mu}(\ddot{\lambda} +
(\dot{\lambda}+\frac{1}{t})(\dot{\lambda}-\dot{\mu})) = 4\pi q
\end{equation}
\begin{equation} \label{eq:1.7}
e^{-2\lambda} \phi'' - e^{-2\mu} \ddot{\phi} -
e^{-2\mu}(\dot{\lambda} - \dot{\mu} + \frac{2}{t})\dot{\phi} -
e^{-2\lambda}(\lambda' - \mu')
 \phi' = 0
\end{equation}
where (\ref{eq:1.7}) is the wave equation in $\phi$ and :
\begin{equation} \label{eq:1.8}
\begin{aligned}
 \rho(t,r)  = e^{-2\mu} T_{00}(t,r) &= \frac{\pi}{t^{2}}
\int_{-\infty}^{+\infty} \int_{0}^{+\infty} \sqrt{1 + w^{2} +
F/t^{2}} f(t,r,w,F)dFdw \\
 & \qquad + \frac{1}{2}(e^{-2\mu}
 \dot{\phi}^{2} +
e^{-2\lambda}{\phi'}^{2})
\end{aligned}
\end{equation}
\begin{equation} \label{eq:1.9}
\begin{aligned}
p(t,r)  = e^{-2\lambda} T_{11}(t,r) &= \frac{\pi}{t^{2}}
\int_{-\infty}^{+\infty}\int_{0}^{+\infty} \frac{w^2}{\sqrt{1
 + w^2 + F/t^{2}}} f(t,r,w,F)dFdw \\
 & \qquad + \frac{1}{2}(e^{-2\mu}
\dot{\phi}^{2} + e^{-2\lambda}{\phi'}^{2})
\end{aligned}
\end{equation}
\begin{equation} \label{eq:1.10}
j(t,r) = -e^{-(\lambda + \mu)} T_{01}(t,r) = \frac{\pi}{t^{2}}
\int_{-\infty}^{+\infty}\int_{0}^{+\infty}wf(t,r,w,F)dFdw
-e^{-(\lambda + \mu)} \dot{\phi} \phi'
\end{equation}
\begin{equation} \label{eq:1.11}
\begin{aligned}
q(t,r) &= \frac{2}{t^{2}} T_{22}(t,r) = \frac{2}{t^{2}\sin_k^2\theta} T_{33}(t,r,\theta)\\
  &= \frac{\pi}{t^{4}} \int
_{-\infty}^{\infty}\int_{0}^{\infty} \frac{F}{\sqrt{1 + w^{2}
 +F/{t^{2}}}} f(t,r,w,F)dFdw + e^{-2\mu} \dot{\phi}^{2} -
e^{-2\lambda}{\phi'}^{2}
\end{aligned}
\end{equation}
We prescribe initial data at time $t=1$:
\begin{eqnarray}
&&f(1,r,w,F) = \overset{\circ}{f}(r,w,F),\ \  \lambda(1,r) =
\overset{\circ}{\lambda}(r),\ \  \mu(1,r) =
\overset{\circ}{\mu}(r),\nonumber      \\
&&\phi(1,r) = \overset{\circ}{\phi}(r),\ \ \dot{\phi}(1,r) =
\psi(r)\nonumber
\end{eqnarray}
The choice $t=1$ is made only for convenience. Analogous results hold
in the case of prescribed data on any hypersurface $t=t_0>0$.

The paper is organized as follows. In section $2$, we show that
the solution of the Cauchy problem corresponding to system
(\ref{eq:1.2})-(\ref{eq:1.11}) exists for all $t \ge 1$ and $k \le
0$.  In section $3$, we prove that the spacetime obtained is
timelike and null geodesically complete towards the future in the
plane symmetric case with $f$  identically zero. This section ends
with results on the homogeneous models. We prove in section $4$, for plane symmetric solutions
with $f=0$, that
the maximal globally hyperbolic development of any initial data is inextendible.

Unless otherwise specified in what follows, constants denoted by C will be positive, may depend on the
 initial data and may change their value from line to line.

\section{Global existence in the future}

 In this section we make use of the continuation criterion in the following local existence result.
\begin{theorem} \label{T:1}
Let
 $\overset{\circ}{f} \in C^{1}(\mathbb{R}^{2} \times [0, \infty[)$
with \\ $\overset{\circ}{f}(r+1,w,F) = \overset{\circ}{f}(r,w,F)$
for $(r,w,F) \in \mathbb{R}^{2} \times [0, \infty[$,
 $\overset{\circ}{f}\geq 0$, and
\begin{eqnarray*}
w_{0} := \sup \{ |w| | (r,w,F) \in {\rm supp} \overset{\circ}{f}
\} <
 \infty
\end{eqnarray*}
\begin{eqnarray*}
F_{0} := \sup \{ F | (r,w,F) \in {\rm supp} \overset{\circ}{f} \}
< \infty
\end{eqnarray*}
Let $\overset{\circ}{\lambda}, \psi \in C^{1}(\mathbb{R})$,
$\overset{\circ}{\mu}, \overset{\circ}{\phi} \in
C^{2}(\mathbb{R})$ with $\overset{\circ}{\lambda}(r) =
\overset{\circ}{\lambda}(r+1)$, $\overset{\circ}{\mu}(r) =
\overset{\circ}{\mu}(r+1)$,\\ $\overset{\circ}{\phi}(r) =
\overset{\circ}{\phi}(r+1)$  and
\begin{eqnarray*}
\overset{\circ}{\mu}'(r) =
 -4 \pi e^{\overset{\circ}{\lambda} +
\overset{\circ}{\mu}}\overset{\circ}{j}(r) , \ \  r \in \mathbb{R}
\end{eqnarray*}
Then there exists a unique, right maximal, regular solution $(f,
\lambda, \mu, \phi)$ of system (\ref{eq:1.2})-(\ref{eq:1.11}) with
$(f, \lambda, \mu, \phi)(1) = (\overset{\circ}{f},
\overset{\circ}{\lambda}, \overset{\circ}{\mu},
\overset{\circ}{\phi})$ and $\dot{\phi}(1) = \psi$ on a time
interval $[1,T[$ with $T \in [1,\infty]$. If
\begin{eqnarray*}
\sup \{ \mid w \mid | (t, r, w, F) \in {\rm supp} f \} < \infty
\end{eqnarray*}
and
\begin{eqnarray*}
\sup \{ e^{2\mu(t, r)} | r \in \mathbb{R}, \ t\in [1, T[ \} <
\infty
\end{eqnarray*}
 then $T = \infty$.
\end{theorem}
This is the content of theorems $4.6$ and $4.7$ in \cite{tegankong}. For a regular solution,
 all derivatives which appear in the system exist and are continuous by definition (see \cite{tegankong}).
 We prove the following result in the cases of plane and
 hyperbolic symmetries.
 \begin{proposition}\label{P:1}
Assume that $(f,\lambda,\mu,\phi)$ is a right maximal regular
solution of the Einstein-Vlasov-scalar field system
with initial data given for $t=1$. Then for $k=0$ or $k=-1$,
 \begin{eqnarray*}
\sup \{e^{2\mu(t, r)} | r \in \mathbb{R}, \ t\in [1, T[ \} <
\infty .
\end{eqnarray*}
\end{proposition}
{\bf Proof} : We now establish a series of estimates which will
result in an upper bound on $\mu$.
 Similar estimates were used in
\cite{tegankong} and \cite{andreasson}.
Firstly, integration of (\ref{eq:1.4}) with respect to $t$ and the fact that
$p$ is non-negative imply that
\begin{equation}\label{eq:2.1}
e^{2 \mu(t, r)} = \left[\frac{e^{-2 \overset{\circ}{\mu}(r)} +
k}{t} - k - \frac{8 \pi}{t}\int_{1}^{t}s^{2}p(s, r)
ds\right]^{-1} \ \geq \frac{t}{C_0-kt}, \ t \in [1, T[
\end{equation}
where $C_0 = k+\sup\{e^{-2 \overset{\circ}{\mu}(r)}; r\in [0,1]\}$.
Next let
 us show that
\begin{equation}\label{eq:2.2}
\int_{0}^{1} e^{\mu+\lambda}\rho(t,r) dr \leq C t, \ \  t \in [1, T[
\end{equation}
A lengthy calculation shows that, since $\mu$ and $j$ are periodic with respect to $r$ :
\begin{equation}\label{eq:2.3}
\frac{d}{dt}\int_{0}^{1} e^{\mu+\lambda}\rho(t,r) dr =
-\frac{1}{t}\int_{0}^{1}e^{\mu+\lambda}\left[2\rho+q
-\frac{\rho+p}{2}(1+ke^{2\mu})\right]dr
\end{equation}
For $k=0$, since $q \geq -2\rho$ and $p\le \rho$, we have:
\begin{equation*}
\frac{d}{dt}\int_{0}^{1} e^{\mu+\lambda}\rho(t,r) dr \leq
\frac{1}{t}\int_{0}^{1} e^{\mu+\lambda}\rho(t,r) dr
\end{equation*}
Integrating with respect to $t$ yields
(\ref{eq:2.2}) for $k=0$. For $k=-1$, we have, using (\ref{eq:2.1}) and the fact that
 $2\rho+q \geq -\rho$, $ p\le \rho $:
\begin{align*}
 \frac{d}{dt}\int_{0}^{1} e^{\mu+\lambda}\rho(t,r) dr &= -\frac{1}{t}\int_{0}^{1}e^{\mu+\lambda}\left[2\rho+q
-\frac{\rho+p}{2}(1-e^{2\mu})\right]dr\\
& \leq -\frac{1}{t}\int_{0}^{1}e^{\mu+\lambda}(2\rho+q
-\frac{\rho+p}{2})dr -\frac{1}{C_0+t}\int_{0}^{1}e^{\mu+\lambda}
\frac{\rho+p}{2}dr\\
& \leq (\frac{2}{t} -\frac{1}{C_0+t})\int_{0}^{1}e^{\mu+\lambda}
\rho dr
\end{align*}
 Integrating with
respect to $t$ yields (\ref{eq:2.2}) for $k=-1$.
 Using (\ref{eq:1.5}), the fact that $|j| \leq \rho$
and (\ref{eq:2.2}) we find
\begin{align*}
\mid \mu(t,r)- \int_{0}^{1}\mu(t,\sigma)d\sigma \mid & =
 \mid \int_{0}^{1}\int_{\sigma}^{r}\mu'(t,\tau)d\tau d\sigma \mid
 \leq \int_{0}^{1}\int_{0}^{1}|\mu'(t,\tau)|d\tau d\sigma \\
 &
 \leq 4\pi t\int_{0}^{1}e^{\mu+\lambda}\rho(t,\tau)d\tau
\end{align*}
that is
\begin{equation}\label{eq:2.4}
  \mid \mu(t,r)- \int_{0}^{1}\mu(t,\sigma)d\sigma \mid
  \leq Ct^{2}, \ t \in [1, T[, \ r \in [0,1]
\end{equation}
 Next we show that
\begin{equation}\label{eq:2.5}
  e^{\mu(t,r)-\lambda(t,r)} \leq Ct , \ t \in [1, T[, \ r
  \in [0,1].
\end{equation}
By (\ref{eq:2.1}) and since $\rho\ge p$,
\begin{align*}
 \frac{ \partial}{\partial t}e^{\mu-\lambda} & =
 e^{\mu-\lambda}\left[4\pi
te^{2\mu}(p-\rho)+\frac{1+ke^{2\mu}}{t}\right] \leq (\frac{1}{t}+\frac{k}{C_0-kt})e^{\mu-\lambda};
\end{align*}
 and integrating this inequality with
respect to $t$ yields
\begin{equation*}
e^{\mu-\lambda} \leq C \frac{t}{C_0-kt} \leq Ct,
\end{equation*}
i.e. (\ref{eq:2.5}).

 Now we estimate the average of $\mu(t)$ over the
interval $[0,1]$ which in combination with (\ref{eq:2.4}) will yield
the desired upper bound on $\mu$. We use (\ref{eq:1.4}), (\ref{eq:2.2}),
(\ref{eq:2.5}) and the fact that $p \le \rho$,  $ke^{2\mu} \le 0$ :
\begin{align*}
\int_{0}^{1}\mu(t,r) dr & = \int_{0}^{1}\overset{\circ}{\mu}(r) dr
+ \int_{1}^{t}\int_{0}^{1}\dot{\mu}(s,r) dr ds\\
& \leq \int_{0}^{1}\overset{\circ}{\mu}(r) dr + \frac{1}{2}\ln t+
4\pi\int_{1}^{t}\int_{0}^{1}se^{\mu-\lambda}e^{\mu+\lambda}\rho dr
ds
+\int_{1}^{t}\int_{0}^{1}\frac{ke^{2\mu} }{2s} dr ds\\
& \le \int_{0}^{1}\overset{\circ}{\mu}(r) dr  + \frac{1}{2}\ln t + Ct^4
\end{align*}
 with (\ref{eq:2.4}) this implies
\begin{equation}\label{eq:2.6}
\mu(t,r) \leq C(1+  \ln t + t^{4}+t^2) \leq Ct^4, \ t \in [1, T[,
\ r \in [0,1].
\end{equation}
\begin{remark}\label{R:1}
We have proven that for initial data as in theorem \ref{T:1}, the
right maximal regular solution of the full system  satisfies
estimates (\ref{eq:2.2}), (\ref{eq:2.5}) and (\ref{eq:2.6}).
\end{remark}
In the next theorem, we prove that the solution exists on the
full interval $[1, \infty[$.

\begin{theorem}\label{T:2}
Assume that $(f,\lambda,\mu,\phi)$ is a solution of the full
system on a right maximal interval of existence $[1,T[$ with initial data as in theorem \ref{T:1}, then
$T=\infty$.
\end{theorem}
{\bf Proof}: Assume that $T < \infty$. We show that under the
bound of $\mu$, we obtain the bound of \ \  $\sup \{ |w||(t,r,w,F)
\in {\rm supp}f\}$ \ \   which is in contradiction to theorem \ref{T:1}.
The proof of the bound on $w$ is similar to the proof of
[see \cite{rein1}, theorem $6.2$].\\
Except in the vacuum case we have $w_0 >0$ and $F_0 >0$. For $ t
\geq 1$ define
\begin{eqnarray*}
 P_{+}(t) := \max \{0, \max \{w | (r,w,F) \in {\rm supp} f(t) \} \},
\end{eqnarray*}
\begin{eqnarray*}
 P_{-}(t) := \min \{0, \min \{ w | (r,w,F) \in {\rm supp} f(t) \}
 \}.
\end{eqnarray*}
Let $(r(s), w(s), F)$ be a characteristic in the support of $f$.
Assume that $P_{+}(t) > 0$ for some $t \in [1,T[$, and let
$w(t)>0$. We have
\begin{align}\label{eq:2.7}
 \dot{w}(s) & =
 -\dot{\lambda}w-e^{\mu-\lambda}\mu' \sqrt{1+w^2+F/s^2}\nonumber\\
 &= \frac{4
\pi^{2}}{s}e^{2\mu}\int_{-\infty}^{\infty}\int_{0}^{\infty}
\left(\tilde{w}\sqrt{1+w^{2}+F/s^{2}}-w\sqrt{1+\tilde{w}^{2}+\tilde{F}/s^{2}}\right)f
d\tilde{F}d\tilde{w} \nonumber\\
& +\frac{1+ke^{2\mu}}{2s}w - 4\pi
se^{2\mu}\left[\frac{1}{2}w(\dot{\phi}^2e^{-2\mu}+\phi'^2e^{-2\lambda})+
\sqrt{1+w^{2}+F/s^{2}}\dot{\phi}\phi'e^{-\mu-\lambda} \right]
\end{align}
Set $\gamma =
\tilde{w}\sqrt{1+w^{2}+F/s^{2}}-w\sqrt{1+\tilde{w}^{2}+\tilde{F}/s^{2}}$.
As long as $w(s) >0$, we have the following estimates: if
$\tilde{w} \leq 0$ then $\gamma \leq 0$. If $\tilde{w} >0$ then
\begin{align}\label{eq:2.8}
 &\frac{4\pi^{2}}{s}e^{2\mu}\int_{-\infty}^{\infty}\int_{0}^{\infty}
\gamma f
d\tilde{F}d\tilde{w} + \frac{1+ke^{2\mu}}{2s}w \leq \frac{1+ke^{2\mu}}{2s}w \nonumber\\
& +
\frac{4\pi^{2}}{s}e^{2\mu}\int_{0}^{P_{+}(s)}\int_{0}^{F_{0}}\frac{\tilde{w}^{2}
(1+w^{2}+F/s^{2})-w^{2}(1+\tilde{w}^{2}+\tilde{F}/s^{2})}{\tilde{w}
\sqrt{1+w^{2}+F/s^{2}}+  w\sqrt{1+\tilde{w}^{2}+\tilde{F}/s^{2}}}f
d\tilde{F}d\tilde{w} \nonumber\\
& \leq
\frac{4\pi^{2}}{s}e^{2\mu}\int_{0}^{P_{+}(s)}\int_{0}^{F_{0}}\frac{\tilde{w}(1+F)}{w}f
d\tilde{F}d\tilde{w}+\frac{1+ke^{2\mu}}{2s}w \nonumber\\
& \leq 4 \pi^{2}F_0 (1+F_0)\parallel
\overset{\circ}{f}\parallel\frac{e^{2\mu}}{2s}(P_{+}(s))^{2}\frac{1}{w}+
\frac{1+ke^{2\mu}}{2s}w \nonumber\\
& \leq \frac{C}{s}\left[\frac{(P_{+}(s))^{2}}{w}+ w\right]
\end{align}
It is proved in \cite{tegankong} that $K(s) = \sup\{(e^{-2\mu}\dot\phi^2+e^{-2\lambda}\phi'^2)(s,r);s\in [1,T[\}
 \leq K(1) s$, then
\begin{align}\label{eq:2.9}
&\frac{1}{2}w(\dot{\phi}^2e^{-2\mu}+\phi'^2e^{-2\lambda})+
\sqrt{1+w^{2}+F/s^{2}}\dot{\phi}\phi'e^{-\mu-\lambda}\nonumber\\
&\geq (w+\sqrt{1+w^{2}+F/s^{2}})\dot{\phi}\phi'e^{-\mu-\lambda}
\nonumber\\
& \geq -\sqrt{1+w^{2}+F/s^{2}}K(s) \nonumber\\
& \geq -C(1+P_{+}(s))s
\end{align}
Therefore, using (\ref{eq:2.8}) and (\ref{eq:2.9}), (\ref{eq:2.7})
yields:
\begin{equation*}
\dot{w}(s) \leq \frac{C}{s}\left[\frac{(P_{+}(s))^{2}}{w(s)}+
w(s)\right] + C(1+P_{+}(s))s
\end{equation*}
i.e.
\begin{equation*}
\frac{d}{ds}w(s)^2 \leq C(s^{-1}+ s)(P_{+}(s))^{2}+
CP_{+}(s)s
\end{equation*}
as long as $w(s) >0$. Let $t_1 \in [1,t[$ be defined minimal such
that $w(s) >0$ for $s \in [t_1,t[$, then
\begin{equation*}
w(t)^2 \leq w(t_1)^2 + C\int_{t_{1}}^t[(s^{-1}+
s)(P_{+}(s))^{2}+ s P_{+}(s)]ds
\end{equation*}
If  $t_1=1$ then $w(t_1) \leq
w_0$ and
\begin{equation*}
w(t)^2 \leq w_0^2 + C\int_1^t[(s^{-1}+
s)(P_{+}(s))^{2}+ sP_{+}(s)]ds.
\end{equation*}
If  $t_1>1$ then $w(t_1)=0$ and
\begin{align*}
w(t)^2 &\leq  C\int_{t_{1}}^t[(s^{-1}+
s)(P_{+}(s))^{2}+ s P_{+}(s)]ds\\
 & \leq w_0^2 + C\int_1^t[(s^{-1}+
s)(P_{+}(s))^{2}+ sP_{+}(s)]ds.
\end{align*}
Thus
\begin{align*}
(P_{+}(t))^2 &\leq w_0^2 + C\int_1^t[(s^{-1}+
s)(P_{+}(s))^{2} + sP_{+}(s)]ds \\
 &\leq w_0^2 + C\int_1^t[(s^{-1}+
s)(P_{+}(s))^{2}+ s]ds \\
& \leq (w_0^2 + Ct^{2})+ C\int_1^t(s^{-1}+
s)(P_{+}(s))^{2}ds ; \ \textrm{for all $t \in [1,T[$}
\end{align*}
 Applying Gronwall's inequality to this estimate
implies that $P_+$ is bounded on $[1,T[$.\\
Estimating $\dot{w}(s)$ from below in the case $w(s)<0$ along the
same lines shows that $P_-$ is bounded as well and the proof is
complete.
\begin{remark}
In the spherically symmetric case ($k=1$), there is no global
existence in the future, regardless of the size of initial data.
For any solution ($f,\lambda,\mu,\phi$), the estimate
\begin{equation*}
e^{-2\mu(t, r)} = \frac{e^{-2 \overset{\circ}{\mu}(r)} +
1}{t} - 1 - \frac{8 \pi}{t}\int_{1}^{t}s^{2}p(s, r)ds \le \frac{e^{-2 \overset{\circ}{\mu}(r)} + 1}{t} - 1
\end{equation*}
has to hold on the interval $[1,T[$. Since the right hand side of
this inequality tends to $-1$ for $t \to \infty$, it follows that
$T<\infty$. Furthermore we deduce from the previous theorem that
$||e^{2\mu}|| \to \infty$ for $t \to T$.
\end{remark}

 \section{The future asymptotic behaviour}
In this section we prove that the spacetime
obtained in theorem \ref{T:2} (with $f=0$) in the plane symmetric case is timelike and
null geodesically complete in the expanding direction. Later on, we prove that this result
holds for homogeneous solutions of the Einstein-Vlasov-scalar field system in plane and hyperbolic symmetries.\\

 \subsection{Particular explicit solutions}
\ \  Let us determine first the explicit solutions $\phi,\mu,\lambda$ in the case $f=0$ and $k=0$.

  In the case $k=0$ the wave equation can be reduced to a simple linear
equation and the result follows from \cite{rendall95}. This reduction goes as follows.
In that case the field equations imply
that $\lambda-\mu+\log t$ is constant in time. It may, however,
be dependent on $r$. Suppose that $r$ is replaced by a new
coordinate $s$ on the initial hypersurface. Choosing $s$
appropriately makes the transformed quantity  $\lambda-\mu+\log t$
constant on the initial hypersurface and hence everywhere.
Once this transformation has been carried out the wave equation
simplifies to $\ddot\phi+t^{-1}\dot\phi=\phi''$.
 We can use the results of Jurke \cite{jurke} obtained in the case of polarized Gowdy $T^3$-models
 (where $W$ is replaced here by $\phi$, $t$ by a multiple of $t^2$ and it is necessary to be careful to get
 the right periodicity of the transformed solution). The equations are similar with the same boundary conditions.
 Following this, for all $(t,r) \in [1,\infty[ \times \mathbb{R}$,  $\phi$ must be in this form :
 \begin{equation}\label{eq:2.10}
\phi(t,r) =
\begin{cases}
a_1 + 2b\log t \ \  &\text{(homogeneous case)}\\
a_1 + 2b\log t + t^{-1}\alpha(t,r) + \beta(t,r) \ \ &\text{(non-homogeneous case)}
\end{cases}
\end{equation}
where $a_1$ and $b$ are constants fixed by initial values of $\phi$ and $\dot\phi$,
$\alpha$ and $\beta$ real-valued $C^2$-functions with
\ \ $|\alpha|,\ |\alpha'| \le C$,\ \ $|\dot{\alpha}| \le Ct$, \ \ $|\beta|,\ |\beta'| \le Ct^{-3}$,\ \
$|\dot{\beta}| \le Ct^{-2}$, \ \ and \ \ $\frac{1}{4}(t^{-2}\partial_{tt} - t^{-3}\partial_t)\alpha(t,r) = \partial_{rr}\alpha(t,r)$. With this, $|\dot{\phi}\phi'| \le Ct^{-1}$.\\
From the field equations (\ref{eq:1.3}), (\ref{eq:1.4}) we have :
\[
\dot\lambda =  -\frac{1}{2t}+ 2\pi t (\dot{\phi}^2+ 4t^2\phi'^2)\
\ \textrm{and}\ \ \dot\mu =  \frac{1}{2t}+ 2\pi t (\dot{\phi}^2+
4t^2\phi'^2)
\]
 Therefore $\dot\mu$ and $\dot\lambda$ are bounded each by a positive constant $C$.
 From Theorem $15$ of \cite{jurke} (the hypothesis of this theorem are satisfied :
 $a_t$ is replaced  here by $\dot\mu$),  $\mu$ can be cast for all $(t,r) \in [1,\infty[ \times \mathbb {R}$
 into the form
\begin{equation}\label{eq:2.11}
\mu(t,r)=
\begin{cases}
\frac{1}{2}(16\pi b^2+1)\log t + \gamma \ \  &\text{if}\ \  \mu'=0 \\
\nu t^2 + \delta(t,r) \ \ &\text{(non-homogeneous case)}
\end{cases}
\end{equation}
where $\nu$ is a positive constant, $\gamma$ a constant fixed by initial value of $\mu$
and the function $\delta$ satisfies the inequalities, \ \ $|\delta(t,r)| \le C(1+t)$,\ \  $|\dot\delta| \le Ct$.
 Note that if $\phi'=0$ then from equation $(1.5)$, $\mu'=0$ and $\lambda'=0$;
 the solutions are independent of $r$ and we are in the homogeneous case.

\subsection{Geodesic completeness}
Let $]\tau_{-}, \tau_{+}[ \ni
\tau \mapsto (x^{\alpha}(\tau), p^{\beta}(\tau))$ be a geodesic
whose existence interval is maximally extended and such that
$x^{0}(\tau_{0}) = t(\tau_{0}) = 1 $ for some $\tau_{0} \in
]\tau_{-}, \tau_{+}[$. We want to show that for future-directed
timelike and null geodesics, $\tau_{+} = + \infty$. Consider first
the case of a timelike geodesic, i.e.,
\begin{equation*}
g_{\alpha\beta}p^{\alpha}p^{\beta} = -m^2 \ ; \ p^{0}>0
\end{equation*}
with $m>0$. Since \ $ dt/d\tau = p^{0}>0$, the geodesic can
be parameterized by the coordinate time $t$. Recall that along the geodesics the variables \\
$t$,
$r$, $p^{0}$, $w:=e^{\lambda}p^{1}$,
$F:=t^{4}\left[(p^{2})^{2}+\sin_{k}^{2}\theta(p^{3})^{2}\right]$
satisfy the following system of differential equations :
\begin{equation}\label{eq:2.12}
\frac{dr}{d\tau} = e^{-\lambda}w, \ \  \frac{dw}{d\tau} =
-\dot{\lambda}p^{0}w - e^{2\mu-\lambda}\mu'(p^{0})^{2}, \ \
\frac{dF}{d\tau} = 0
\end{equation}
\begin{equation}\label{eq:2.13}
\frac{dt}{d\tau} = p^{0}, \ \frac{dp^{0}}{d\tau} =
-\dot{\mu}(p^{0})^{2} -
2e^{-\lambda}\mu'p^{0}w-e^{-2\mu}\dot{\lambda}w^{2}-e^{-2\mu}t^{-3}F.
\end{equation}
With respect to
coordinate time the geodesic exists on the interval $[1,
\infty[$ since on bounded $t$-intervals the Christoffel symbols
are bounded and the right hand sides of the geodesic equations
written in coordinate time are linearly bounded in $p^{1}$, $
p^{2}$, $p^{3}$.
Along the geodesic we define $w$ and $F$ as above. Then the
relation between coordinate time and proper time along the
geodesic is given by
\begin{equation*}
\frac{dt}{d\tau} = p^{0} = e^{-\mu}\sqrt{m^{2}+w^{2}+F/t^{2}},
\end{equation*}
and to control this we need to control $w$ as a function of
coordinate time.

\subsubsection{The plane symmetric case without Vlasov}

Assume that $w(t)>0$ for some $t \ge 1$.  By (\ref{eq:2.12}), the fact that
$\ |\dot\phi||\phi'|(s)\le Cs^{-1}$, $e^{(\mu-\lambda)(s)} = Cs$ and $|j| \le \rho$, we have as long as $w(s)>0$
\begin{equation}\label{eq:2.14}
\begin{aligned}
\frac{d}{ds}w(s) &=
 4\pi se^{2\mu}(j\sqrt{m^{2}+w^2+F/s^2}-\rho w)+\frac{1}{2s}w \\
&\le \frac{1}{2s}w -4\pi se^{2\mu}\rho w +4\pi se^{2\mu}|j|(w + \sqrt{m^{2}+F/s^2})\\
&\le \frac{1}{2s}w  + 4\pi se^{\mu-\lambda}|\dot\phi||\phi'|\sqrt{m^{2}+F/s^2}\\
& \le \frac{1}{2s}w  + Cs
\end{aligned}
\end{equation}
Let $t_0 \in [1,t[$ be defined minimal such that $w(s)>0$ for $s \in [t_0,t[$. Then
Gronwall's inequality shows that
\begin{equation*}
w(t) \le \left[ w(t_0) +C\int_{t_0}^t s \exp(\int_s^{t_0} \frac{d\tau}{2\tau})ds\right]
\exp(\int_{t_0}^t \frac{d\tau}{2\tau}).
\end{equation*}
Now either $t_0 = 1$ and $w(t_0) = w(1)$ or $t_0 >1$ and $w(t_0) = 0$. Thus
\begin{equation*}
w(t) \le \left[ |w(1)| +C\int_1^t s \exp(\int_s^1 \frac{d\tau}{2\tau})ds\right]
\exp(\int_1^t \frac{d\tau}{2\tau})  \le Ct^{2}.
\end{equation*}
Estimating $\dot{w}(s)$ from below in the case $w(s) < 0$ along the same lines yields the upper bound of $-w(t)$.\\
 Since $|\delta| \le C(1+t)$,\  $\nu t^2 +\delta > \nu t^2(1-\frac{C(1+t)}{\nu t^2})$.
 For $t$ large, we choose $\frac{C(1+t)}{\nu t^2} <\frac{1}{2}$.
Then along the geodesic we have :
\begin{equation*}
\frac{d\tau}{dt} = \frac{e^{\mu}}{\sqrt{m^{2}+w^{2}+F/t^{2}}}  \geq
\frac{\nu t^2}{2t^{2}\sqrt{m^{2}+C+F}}.
\end{equation*}
  Since the left hand side is constant, the integral over $[1, \infty[$
diverges.\\
In the homogeneous case ($\nu =0$, $j=0$, $\rho \ge 0$), (\ref{eq:2.14}) becomes
\begin{equation*}
\frac{d}{ds}w(s) \le \frac{1}{2s}w
\end{equation*}
i.e \  $w(t) \le Ct^{1/2}$ for $t \ge 1$. Therefore
\begin{equation*}
\frac{d\tau}{dt}  \ge Ct^{8\pi b^2}
\end{equation*}
The integral of the right hand side of the above inequality over $[1, \infty[$ diverges.\\
 In either case, we conclude that $\tau_{+} = + \infty$ as desired.\\
In the case of a
future-directed null geodesic, i.e. $m=0$ and $p^{0}(\tau_{0})>0$,
$p^0$ is everywhere positive and the quantity $F$ is again
conserved. The argument can now be carried out exactly as in the
timelike case, implying that $\tau_{+}= +\infty$. We have proven :

\begin{theorem}\label{T:3}
Consider initial data with plane
symmetry for the Einstein-scalar field system written in areal coordinates. Then the
corresponding spacetime is timelike and null geodesically complete in the expanding direction.
\end{theorem}

\subsubsection{The spatially homogeneous case}
Consider now the Einstein-Vlasov-scalar field system for $k \le 0$.
We study the future behaviour for solutions which are independent of the space coordinate $r$. In this case $\mu' =0$, $\phi'=0$ and equation (\ref{eq:1.5}) shows that $j=0$.

We prove that $w(t)$ is bounded for finite time $t$ and the spacetime obtained is geodesically complete.
 Since $\rho \ge 0$, $k \le 0$, we have from (\ref{eq:2.12}),
\begin{align*}
\frac{d}{ds}w(s) = -4\pi se^{2\mu}\rho w+\frac{1+ke^{2\mu}}{2s}w \le \frac{1}{2s}w
\end{align*}
and we can prove as above that $w(t)\le Ct^{1/2}$ for $t\ge 1$. Using (\ref{eq:2.1}), along the geodesic,
\begin{align*}
\frac{d\tau}{dt} \ge \frac{C}{\sqrt{m^{2}+Ct+F/t^{2}}}  \geq C t^{-1/2}.
\end{align*}
The integral of the right hand side over $[1, \infty[$
diverges. Therefore $\tau_+= +\infty$. In the case of a future-directed null geodesic, the argument is the same as what we have done in the previous subsection.

In the homogeneous case, a number of further estimates can also be obtained.
 We follow the proof of theorem \ref{T:1} and \cite{rein2} to obtain a better bound of $\mu$ which depend explicitly of the data. Recall (\ref{eq:2.3})
\begin{equation*}
\frac{d}{dt} (e^{\mu+\lambda}\rho)(t) =-\frac{1}{t}e^{\mu+\lambda}\left[2\rho+q
-\frac{\rho+p}{2}(1+ke^{2\mu})\right].
\end{equation*}
Since $q\ge 0$, we obtain :\\
for $k = 0$,
\begin{equation*}
\frac{d}{dt} (e^{\mu+\lambda}\rho)(t) \leq -\frac{1}{t}e^{\mu+\lambda}\rho(t);
\end{equation*}
thus
\begin{equation}\label{eq:2.15}
e^{\mu+\lambda}\rho(t) \leq e^{\overset{\circ}{\mu}+\overset{\circ}{\lambda}} \overset{\circ}{\rho}t^{-1};
\end{equation}
for $k=-1$
\begin{align*}
\frac{d}{dt} (e^{\mu+\lambda}\rho)(t) &\leq -\frac{1}{t}e^{\mu+\lambda}\left[2\rho
-\frac{\rho+p}{2}-\frac{1}{C_0+t}\frac{\rho+p}{2}\right]\\
& \le e^{\mu+\lambda}\rho\left(-\frac{1}{t}-\frac{1}{C_0+t}\right)\\
& \le -\frac{2}{C_1+t}e^{\mu+\lambda}\rho(t)
\end{align*}
where $C_0 = k+ e^{-2\overset{\circ}{\mu}}$ and $C_1 = \max (0, C_0)$.
Thus
\begin{equation}\label{eq:2.16}
e^{\mu+\lambda}\rho(t) \leq (C_1+1)^2e^{\overset{\circ}{\mu}+\overset{\circ}{\lambda}} \overset{\circ}{\rho}t^{-2}.
\end{equation}
Next we have similarly to (\ref{eq:2.5}),
\begin{equation}\label{eq:2.17}
 e^{\mu-\lambda} \le e^{\overset{\circ}{\mu}-\overset{\circ}{\lambda}}t \ \ \rm{for}\ \ k=0
\end{equation}
\begin{equation}\label{eq:2.18}
 e^{\mu-\lambda} \le  e^{\overset{\circ}{\mu}-\overset{\circ}{\lambda}}\frac {e^{-2\overset{\circ}{\mu}}t}{e^{-2\overset{\circ}{\mu}}-1+t} \le e^{|\overset{\circ}{\mu}|-\overset{\circ}{\lambda}} \ \ \rm{for}\ \ k=-1
\end{equation}
For $k=0$,
\begin{align*}
\mu(t)
&\leq \overset{\circ}{\mu}+\frac{1}{2}\log t+ 4\pi\int_{1}^{t}se^{2\mu}\rho
ds
= \overset{\circ}{\mu}+\frac{1}{2}\log t+
4\pi\int_{1}^{t}se^{\mu-\lambda}e^{\mu+\lambda}\rho ds\\
 & \leq \overset{\circ}{\mu} + \frac{1}{2}\log t + 4\pi e^{\overset{\circ}{\mu}+\overset{\circ}{\lambda}} \overset{\circ}{\rho}e^{\overset{\circ}{\mu}-\overset{\circ}{\lambda}}\int_{1}^{t}s ds
\end{align*}
i.e
\begin{equation}\label{eq:2.19}
\mu(t) \leq \overset{\circ}{\mu}+ 2\pi e^{2\overset{\circ}{\mu}} \overset{\circ}{\rho}  t^2.
\end{equation}
For $k=-1$,
\begin{align*}
\mu(t)
&\leq \overset{\circ}{\mu}+4\pi\int_{1}^{t}se^{2\mu}\rho ds + \int_{1}^{t}\frac{1}{2s}(1-e^{2\mu}) ds \\
& \leq \overset{\circ}{\mu}+
4\pi\int_{1}^{t}se^{\mu-\lambda}e^{\mu+\lambda}\rho ds + \int_{1}^{t}\frac{1}{2s}(1-\frac{s}{C_0 +s}) ds\\
& \leq \overset{\circ}{\mu}+ \frac{1}{2}\log (C_0+1)+ 4\pi (C_1+1)^2e^{|\overset{\circ}{\mu}|+\overset{\circ}{\mu}} \overset{\circ}{\rho}\int_{1}^{t}s^{-1}ds
\end{align*}
i.e
\begin{equation}\label{eq:2.20}
\mu(t) \leq \overset{\circ}{\mu}+  \frac{1}{2}\log (C_0+1) + 4\pi e^{|\overset{\circ}{\mu}|+\overset{\circ}{\mu}} \overset{\circ}{\rho}(C_1+1)^2 \log t.
\end{equation}
 We have proven :
\begin{theorem}\label{T:4}
Consider spatially homogeneous solutions of the plane and hyperbolic symmetric  Einstein-Vlasov-scalar field system written in areal coordinates and initial data given for $t=1$. Then these solutions exist on the whole interval $[1,\infty[$ and  the
corresponding spacetimes are timelike and null geodesically complete in the expanding direction. Estimates (\ref{eq:2.15})-(\ref{eq:2.16})-(\ref{eq:2.17})-(\ref{eq:2.18})-(\ref{eq:2.19})-(\ref{eq:2.20}) are satisfied.
\end{theorem}
\begin{remark}
In the homogeneous plane symmetric case with only a scalar field,
solving the wave equation, we obtain \ $|\dot{\phi}(t)|$ $=
|\psi|t^{-1}$. From the field equations, $\dot\lambda(t) = (2\pi
\psi^2 -\frac{1}{2})t^{-1}$ and \ $\lambda(t)=
\overset{\circ}{\lambda}+ (2\pi \psi^2 -\frac{1}{2})\log t$\ ;\
$\mu(t)= \overset{\circ}{\mu}+ (2\pi \psi^2 +\frac{1}{2})\log t$ \
; \ $\rho(t) = p(t) = \frac{1}{2}\psi^2e^{-2\overset{\circ}{\mu}}
t^{-3-4\pi \psi^2}$ \ ; \ $j(t)=0$\ ;\ $q(t) = 2 p(t)$.\\
 We can compute the limiting values of the generalized Kasner exponents :\\
 \begin{align*}
&\lim_{t\to \infty} \frac{K_{1}^{1}(t,r)}{K(t,r)} = \lim_{t\to \infty} \frac{t\dot\lambda(t,r)}{t\dot\lambda+2} = \frac{4\pi \psi^2 -1}{4\pi \psi^2 +3} \ ; \\
&\lim_{t\to \infty} \frac{K_{2}^{2}(t,r)}{K(t,r)} = \lim_{t\to \infty}
\frac{K_{3}^{3}(t,r)}{K(t,r)} = \lim_{t\to \infty} \frac{1}{t\dot\lambda+2} =\frac{2}{4\pi \psi^2 +3},
\end{align*}
\end{remark}
%\ \ \ Non-homogeneous case : \ \ $t\dot\lambda = 2\nu t^2 +t\dot{\delta}-1$. If \ $\lim_{t\to \infty} \frac{\dot{\delta}}{t}$ exists, then\\
 %\begin{align*}
%&\lim_{t\to \infty} \frac{K_{1}^{1}(t,r)}{K(t,r)} = \lim_{t\to \infty} \frac{2\nu  +\frac{\dot{\delta}}{t}-\frac{1}{t}}{2\nu +\frac{\dot{\delta}}{t}+\frac{1}{t}} = 1 \ ; \\
%&\lim_{t\to \infty} \frac{K_{2}^{2}(t,r)}{K(t,r)} = \lim_{t\to \infty}
%\frac{K_{3}^{3}(t,r)}{K(t,r)} = \lim_{t\to \infty} \frac{1}{t^2}\frac{1}{2\nu +\frac{\dot{\delta}}{t}+\frac{1}{t}} = 0.
%\end{align*}
\section{Strong cosmic censorship}
Consider the Einstein-scalar field system with plane symmetry and
initial data given at $t=1$. It follows from \cite{rendall95} that
the Kretschmann scalar blows up uniformly as the singularity is
approached. This means that the inhomogeneous spacetime cannot
be extended to the past. It cannot be extended to the future for
any initial data since we have proved geodesic completeness in
theorem \ref{T:3}. Therefore the maximal globally hyperbolic
development of any initial data is inextendible. This means in
conclusion that the strong cosmic censorship holds for this model (see \cite{moncrief}).

The question whether the spacetime is geodesically complete in the
non-homogeneous hyperbolic case remains open.

  \textbf{\textit{Acknowledgements}} : the author thanks A.D. Rendall for helpful comments.
  He also thanks N. Noutchegueme. Support by a research grant from the VolkswagenStiftung, Federal Republic of Germany
  is acknowledged. This work was completed during a visit at the Max-Planck Institute for Gravitational Physics, Golm, Germany.

\end{document}